\documentclass[12pt,a4paper]{article}

\usepackage{graphics}
\usepackage{amsmath}

\title{Limits on the Two Higgs Doublet Model
 from meson decay, mixing and CP violation}
\author{Carlos A. Mar\'{\i}n and B. Hoeneisen}
\date{\small{Universidad San Francisco de Quito \\
9 October 2002}}
\begin{document}
\maketitle

\begin{abstract}
\noindent
We calculate the rate of $\pi^{+}$, $K^{+}$, $D^{+}$ and $B^{+}
\rightarrow
\mu^{+} \nu_{\mu}$ decays, the branching ratio corresponding to $H^{+}
\rightarrow \tau^{+} \nu_{\tau}$, and the box diagrams of $B^{o}
\leftrightarrow
\bar{B}^{o}$, $K^{o} \leftrightarrow \bar{K}^{o}$ and
$D^{o} \leftrightarrow \bar{D}^{o}$ mixing in the Two Higgs Doublet
Model (Model II). Using the experimental data on meson decay rates,
mixing, and CP violation in the $K^o$ and $B^o$ systems
we set competitive upper and lower limits to the parameter
$\tan \beta$ as a function of the mass of the charged Higgs $m_H$.
\end{abstract}
\section{Introduction}
The Standard Model of quarks and leptons is here to stay.
This theory is based on principles: special relativity, locality,
quantum mechanics, local symmetries and renormalizability\cite{Wilczek}.
Therefore the predictions of the Standard Model
\textquotedblleft{are precise and unambiguous, and generally cannot
be modified `a little bit' except in very limited specific ways. This
feature makes the experimental success especially
meaningful, since it becomes hard to imagine that the
theory could be approximately right without
in some sense being exactly right.}"\cite{Wilczek}
Among the extensions of the Standard Model that respect its
principles and symmetries, that are compatible with present data
within a region of parameter space, and are of interest at the
large particle colliders, is the addition of a second doublet of
Higgs fields. Higgs doublets can be added to the Standard
Model without upsetting the Z/W mass ratio; higher dimensional
representations upset this ratio.
A second Higgs doublet could make the three running coupling
constants of the Standard Model meet at the Grand Unified Theory
(GUT) scale.
A second Higgs doublet is necessary in
Supersymmetric extensions of the
Standard Model\cite{Peccei}. In this article we explore the limits that present
data place on the parameters of the Two Higgs Doublet Model
(Model II).\cite{1} In particular we consider meson decay, mixing
and CP violation.

All of our analysis is based on the
\textquotedblleft{tree-level Higgs potential}"\cite{1}. The physical spectrum
of the Two Higgs Doublet Model (Model II)
contains five Higgs bosons: one pseudoscalar $A^{o}$ (CP-odd
scalar), two neutral scalars $H^{o}$ and $h^{o}$ (CP-even scalars), and
two charged scalars $H^{+}$ and $H^{-}$. The masses of the Higgs
bosons, the mixing angle $\alpha$ between the two neutral scalar Higgs
fields, and the ratio of the vacuum expectation values of the two
neutral components of the Higgs doublets, $\tan \beta > 0$, are free
parameters of the theory.
\begin{eqnarray*}
\Phi_1 = \left( \begin{array}{c}
\Phi_{1}^{o*} \\-\Phi_{1}^{-}\end{array} \right),
\qquad
\Phi_2 =
\left(\begin{array}{c} \Phi_{2}^{+} \\
\Phi_{2}^{o}\end{array}\right),
\qquad
\tan\beta \equiv
\frac{\langle \Phi_{2}^{o} \rangle}{\langle \Phi_{1}^{o*} \rangle}.
\end{eqnarray*}
Using the experimental data on meson decay rates, mixing and CP violation
we set limits to the parameter $\tan\beta$ as a function of the mass of the
charged Higgs $m_H$.
This article is an update of \cite{revista_colombiana}. The reason for this
update is that the recent measurements of $\sin(2\beta_{CKM})$ by the B-factories
Belle\cite{Belle} and BaBar\cite{BaBar}
permit us to set more stringent limits on $\tan\beta$.
$\beta_{CKM}$ is an angle of the
\textquotedblleft{unitarity triangle}".\cite{5}

\section{Theory}
Consider the $\left( B^{o} , \bar{B}^{o} \right)$ system. $B^{o}
\leftrightarrow \bar{B}^{o}$ mixing occurs because of the box diagrams
illustrated in Figure \ref{Feynman_diagram1}.
The difference in mass of the two eigenstates
that diagonalize the hamiltonian can be written in the form
\begin{equation}
\Delta m_{B} = \frac{\beta_{B} G_{F}^{2} m_{W}^{2}
f_{B}^{2} m_{B}}
{6 \pi^{2}}
\left\vert \sum_{i,j} \xi_{i} \xi_{j} \left[ S^{WW} - 2
\cot^{2}\beta \cdot S^{HW} + \frac{1}{4} \cot^{4} \beta \cdot S^{HH} \right]
\right\vert.
\label{mixing}
\end{equation}
The functions
\begin{eqnarray*}
S^{WW} \left( x_{W}^{i} , x_{W}^{j} \right),
\textrm{ }
S^{HW}\left(x_{W}^{i}, x_{W}^{j}, x_{H}^{i}, x_{H}^{j},
x_{H}^{W} \right)
\textrm{ and }
S^{HH} \left(x_{H}^{i}, x_{H}^{j}, x_{H}^{W} \right)
\end{eqnarray*}
are obtained from
the box diagrams
and are written in Appendix A. The Feynman
rules for $H^{\pm}$ are listed in Appendix B. We have derived\cite{2}
$S^{WW}$ in agreement with the literature\cite{3}. The derivation of
$S^{HW}$ and $S^{HH}$ is given in \cite{4}. The variables of these
functions are
\begin{eqnarray*}
x_{W}^{i} \equiv
\frac{m_{i}^{2}}{m_{W}^{2}},
\qquad
x_{H}^{i} \equiv
\frac{m_{i}^{2}}{m_{H}^{2}},
\qquad
\textrm{and}
\qquad
x_{H}^{W} \equiv
\frac{m_{W}^{2}}{m_{H}^{2}}
\end{eqnarray*}
where $i = u, c, t$.
$\xi_{i} \equiv V_{ib} V_{id}^{*}$.
The notation for the remaining symbols in (\ref{mixing}) is standard\cite{5}.
To obtain the Standard Model\cite{3}, omit $S^{HW}$ and
$S^{HH}$. $\beta_{B}$ is a factor of order 1. Estimates of  $\beta_{B}$
using \textquotedblleft{vacuum intermediate state insertion}"\cite{3},
\textquotedblleft{PCAC and vacuum
saturation}"\cite{3}, \textquotedblleft{bag model}"\cite{3},
\textquotedblleft{QCD corrections}"\cite{6,7},
and the \textquotedblleft{free particles in a box}"\cite{2}
models span the range $\approx
0.4$ to $\approx 1$. $f_{B}$ is the decay constant that appears in the
decay rate for $B^{+} \rightarrow \mu^{+} \nu_{\mu}$\cite{5} which at
tree level in the Two Higgs Doublet Model (Model II) is:
\begin{figure}
\begin{center}
\scalebox{1}
{\includegraphics{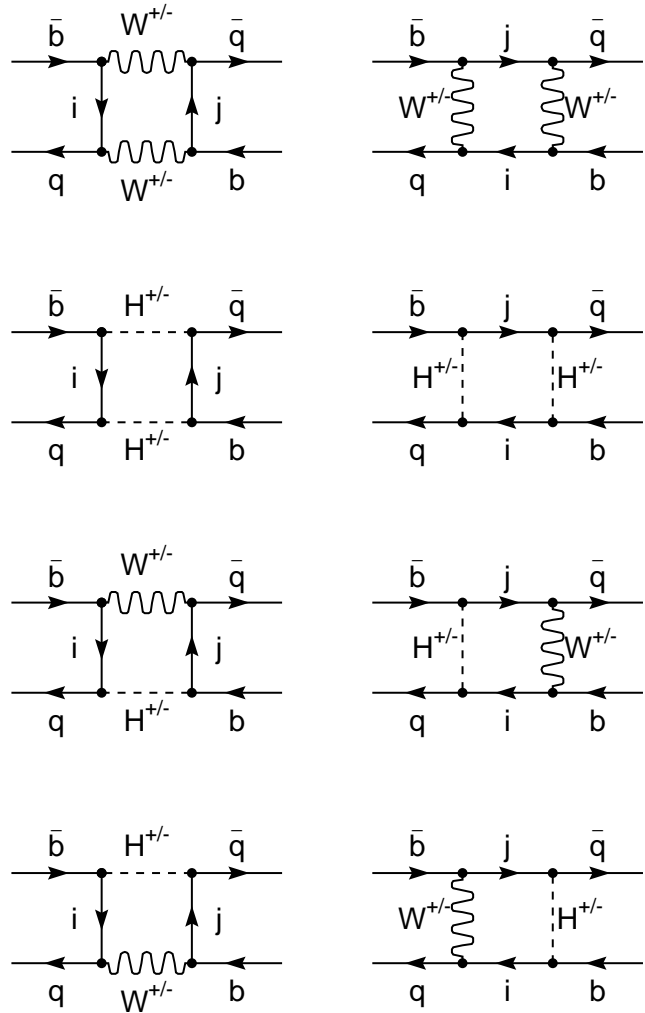}}
\caption{Feynman diagrams corresponding to $B^{o} \leftrightarrow
\bar{B}^{o}$ mixing in the Two Higgs Doublet Model. $q = d$ or $s$ and
$i, j = u, c, t$. The diagrams on the right side interfer with a
\textquotedblleft{-}" sign.}
\label{Feynman_diagram1}
\end{center}
\end{figure}
\begin{equation}
\Gamma_{B^{+}} = \frac {\mathopen{\vert} V_{ub}
\mathclose{\vert}^{2}}{8 \pi} G_{F}^{2} m_{\mu}^{2} m_{B^{+}}
\left(1-\frac{m_{\mu}^{2}}{m_{B^{+}}^{2}}\right)^{2} \left[ f_{B} - g_{B}
\frac{m_{B^{+}}^{2}}{m_{H}^{2}} \tan^{2}\beta \right]^{2}
\label{decay}
\end{equation}
In the derivation of (\ref{decay}) we have substituted
\begin{eqnarray*}
\bar{v}\left(\bar{\textrm{b}}\right) \gamma^{\mu} \left(1 - \gamma^{5}
\right) u \left( \textrm{u} \right) \rightarrow p^{\mu} f_{B}, \\
\bar{v} \left(\bar{\textrm{b}} \right) \left(1 - \gamma^{5}
\right) u \left( \textrm{u} \right) \rightarrow - \frac{m_{B^{+}}^{2}}{m_{b}} g_{B}
\end{eqnarray*}
which defines the decay constants $f_{B}$ and
$g_{B}$. $\bar{v} \left( \bar{\textrm{b}} \right)$ and $u\left(\textrm{u}\right)$ are spinors,
see Appendix B. We expect $f_{B} \approx g_{B}$: for a scalar meson with
the quark and antiquark at rest $f_{B} = \frac{m_{B^{+}}}{m_{b}} g_{B}$.
The decays $B^{+}\rightarrow \mu^{+} \nu_{\mu}$ and $D^{+}\rightarrow
\mu^{+} \nu_{\mu}$ are not yet accessible to experiment so that $f_{B}$
and
$f_{D}$ are unknown. $f_{B}$ is estimated using sum rules\cite{8}, or the
$B^{*} - B$ mass difference\cite{9}, or a phenomenological model\cite{10},
or the MIT bag model\cite{11}. These estimates span the range
$ \approx 0.06 GeV$ to $\approx 0.2 GeV$ with the convention used in
reference \cite{5} and in Equation (\ref{decay}).

In the \textquotedblleft{free particles in a box}"\cite{2} model
$\beta_B = 1$ (after correcting \cite{2} by a color factor $4/3$)
and the volume of the box, \textit{i.e.} the meson, is
$V = 8 / \left( \beta_B m_B f_B^2 \right)$.

For the $\left( B_{s}^{o},
\bar{B}_{s}^{o} \right)$ system: $\xi_{i} \equiv V_{ib}
V_{is}^{*}$ where $ i = u, c, t$; in (\ref{mixing}) replace subscript $B$ by
$B_{s}$. For the $\left( K^{o}, \bar{K}^{o} \right)$
system:
$\xi_{i} \equiv V_{is} V_{id}^{*}$ where $ i = u, c, t$; in (\ref{mixing}) replace
subscript $B$ by $K$. The CP violation parameter $\varepsilon$\cite{3,5}
in the $\left( K^{o}, \bar{K}^{o} \right)$ system in the Two Higgs Doublet
Model is given by:
\begin{equation}
\varepsilon = e^{i \frac{\pi}{4}} \cdot \frac{Im
\left(\sum_{i,j} \xi_{i} \xi_{j} \left[ S^{WW} - 2
\cot^{2}\beta \cdot S^{HW} + \frac{1}{4} \cot^{4} \beta \cdot
S^{HH}\right]\right)}{2 \sqrt{2} \cdot \left\vert \sum_{i,j} \xi_{i}
\xi_{j} \left[ S^{WW} - 2 \cot^{2}\beta \cdot S^{HW} + \frac{1}{4} \cot^{4}
\beta \cdot S^{HH} \right] \right\vert}
\label{epsilon}
\end{equation}
For the $\left( D^{o} , \bar{D}^{o} \right)$ system: $\xi_{i} \equiv V_{ci}
V_{ui}^{*}$ where $i = d, s, b$; in (\ref{mixing}) replace subscript $B$ by $D$
and replace $\cot \beta$ by $\tan \beta$
(leave $\tan \beta$ as is in (\ref{decay})).

The branching ratio for $H^{+} \rightarrow \tau^{+} \nu_{\tau}$
for $m_H < m_t$ is given by
\begin{equation}
B \left( H^{+} \rightarrow \tau^{+} \nu_{\tau} \right)
\approx \\
\frac { m_{\tau}^{2} \tan^{2}\beta}
{\left| V_{cs} \right|^2 a +
\left| V_{cb} \right|^2 b
+ m_\tau^2 \tan^2 \beta }
\label{BR}
\end{equation}
with $a \equiv 3 \left[ m_s^2 \tan^2 \beta + m_c^2 \cot^2 \beta \right]$ and
$b \equiv 3 \left[ m_b^2 \tan^2 \beta + m_c^2 \cot^2 \beta \right]$.
From the measured limit\cite{ALEPH} on $m_H$ as a function
of the branching ratio and (\ref{BR}) we obtain a lower
bound of $m_H$ for each $\tan\beta$.

Let us finally mention that the time-dependent CP-violating
asymmetry $A \equiv (\Gamma - \bar{\Gamma})/(\Gamma + \bar{\Gamma})$,
where $\Gamma$ ($\bar{\Gamma}$) is the rate of the decay
$B^o \rightarrow J/\psi + K_s$ ($\bar{B^o} \rightarrow J/\psi + K_s$),
measured by CDF, Belle and BaBar is given by
$\sin(2\beta_{CKM}) \cdot \sin(\Delta M t)$ in both the
Standard Model and in the Two Higgs Doublet Model (Model II).
This is because the dominating terms of
$\xi_i \xi_j S^{HW}$ and $\xi_i \xi_j S^{HH}$
have $i = j = t$.

\section{Limits}
All experimental data are taken from \cite{5}.
In order to obtain limits
we assume conservatively $0.4 < \beta_x < 1.8$, and $f_x = g_x$
with $x =B$, $B_s$, $D$, $K$, $\pi$.
These assumptions are not critical since the upper (lower)
limits on $\tan\beta$
depend on terms $\propto \tan^4 \beta$ ($\propto \cot^4 \beta$)
in (\ref{mixing}) or (\ref{decay}).
We take the magnitude of the elements of the CKM matrix
from \cite{5} and leave the phase $\angle V_{ub}$ as a free parameter.
The following calculations are made for each $(m_H, \tan\beta)$.
The measured value of the parameter $\varepsilon$ determines the
phase $\angle V_{ub}$
of the CKM matrix, and hence $\beta_{CKM}$. This phase is required to
be within the experimental bounds:
$0.325 < \tan(\beta_{CKM}) < 0.862$ at $95\%$ confidence.\cite{5}
The measured decay rates $\Gamma_K$ and $\Gamma_\pi$
determine $f_K$ and $f_\pi$ using (\ref{decay}). The experimental
upper bounds on $\Gamma_B$ and $\Gamma_D$ determine
upper bounds on $f_B$ and $f_D$ using (\ref{decay}).
The measured $\Delta m_B$ and $\Delta m_K$ determine
$\beta_B f_B^2$ and $\beta_K f_K^2$ using (\ref{mixing}).
The experimental upper bound on $\Delta m_D$ determines
an upper bound on $\beta_D f_D^2$. The experimental lower
bound on $\Delta m_{Bs}$ determines a lower bound on
$\beta_{Bs} f_{Bs}^2$.
From the preceding information we obtain $\beta_K$ and a lower
bound on $\beta_B$. Then the requirements
$0.4 < \beta_K < 1.8$, $\beta_B < 1.8$ and
$0.325 < \tan(\beta_{CKM}) < 0.862$ place limits on
$\tan\beta$ for each $m_H$ as listed in Table \ref{limits_table}.
The confidence level of these limits is $95\%$.
It turns out that the lower limit on $\tan\beta$ is determined by
the experimental lower limit of $\tan(\beta_{CKM})$, and the upper
limit on $\tan\beta$ is determined by $\beta_B < 1.8$.
\begin{table}
\begin{center}
\begin{tabular}{|l|l|}
\hline
$m_H = 100$GeV &  $1.74$ $ < \tan \beta < 67$ \\
$m_H = 200$GeV &  $1.36$ $ < \tan \beta < 134$ \\
$m_H = 300$GeV &  $1.13$ $ < \tan \beta < 202$ \\
$m_H = 1000$GeV & $0.58$ $ < \tan \beta < 672$ \\
\hline
\end{tabular}
\end{center}
\caption{Limits on $\tan \beta$ for several $m_H$ from
measurements of meson decay, mixing and CP violation.
These limits correspond to $95\%$ confidence.}
\label{limits_table}
\end{table}

\begin{figure}
\begin{center}
\scalebox{0.8}
{\includegraphics{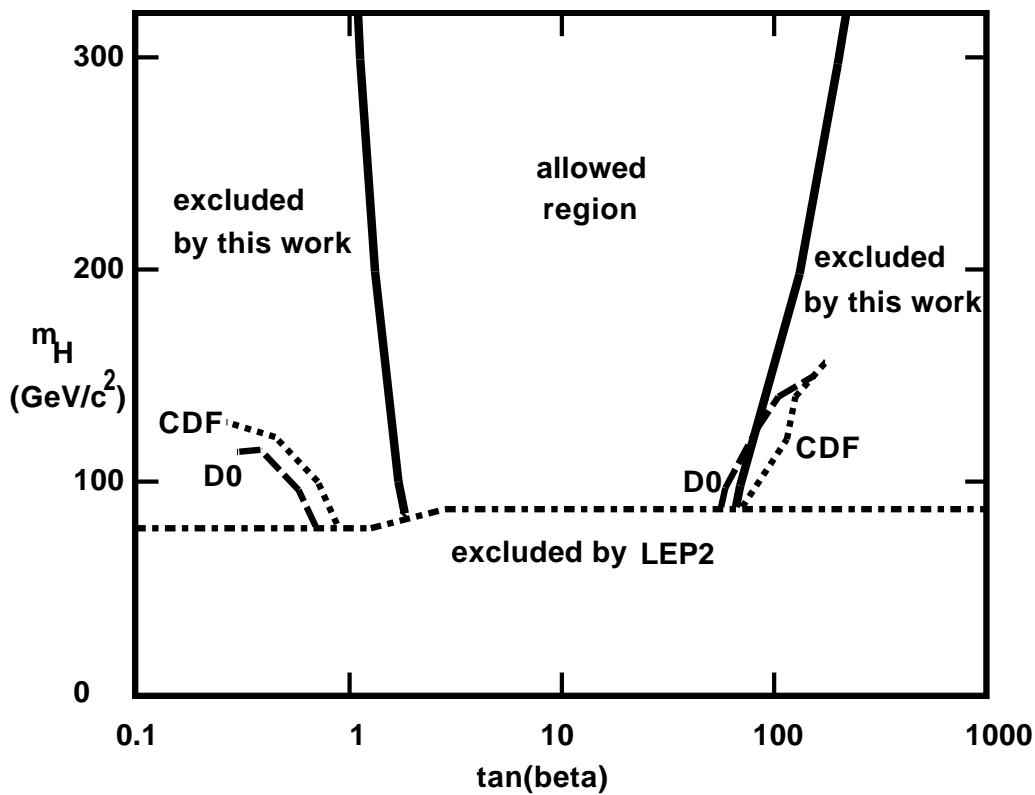}}
\caption{Lower and upper limits on $\tan\beta$ as a
function of the mass of the charged Higgs $m_H$
from meson decay, mixing and CP violation
(continuous curve) compared
to limits obtained by CDF\cite{CDF}, D0\cite{D0} and LEP2\cite{LEP2}, all
at $95\%$ confidence.}
\label{limits_fig}
\end{center}
\end{figure}

\section{Conclusions}
Using measured meson decay rates, mixing and CP violation we
have obtained lower and upper bounds of $\tan\beta$
for each $m_H$. These limits are compared with the
results of direct searches in Figures \ref{limits_fig}.
Note that the measurements of $\sin(2\beta_{CKM})$ by the
Belle and BaBar collaborations have raised the lower bound on $\tan\beta$
by a factor $\approx 5$ with respect to our previous
calculation.\cite{revista_colombiana}

\appendix

\section{Functions $S^{WW}, S^{HW}$ and $S^{HH}$.}

If $i \neq j$:
\begin{eqnarray}
S^{WW}\left(x_{W}^{i}, x_{W}^{j}\right) &=& \frac{x_{W}^{i} + x_{W}^{j}
- \frac{11}{4}
x_{W}^{i}x_{W}^{j}}{\left(1-x_{W}^{i}\right)\left(1-x_{W}^{j}\right)} \nonumber \\
& & + \frac{1}{\left(x_{W}^{i} - x_{W}^{j}\right)}
\left[ G\left(x_{W}^{i},x_{W}^{j} \right) -
G\left(x_{W}^{j},x_{W}^{i}\right)\right]
\label{SWW}
\end{eqnarray}
where
\begin{equation}
G\left(x_{W}^{i},x_{W}^{j}\right) = \frac{\left(x_{W}^{i}\right)^{2}
\ln\left( x_{W}^{i} \right)}{\left( 1 - x_{W}^{i} \right)^{2}} \left[1 - 2
x_{W}^{j} + \frac{1}{4} x_{W}^{i} x_{W}^{j} \right].
\label{G}
\end{equation}
If $i = j$:
\begin{eqnarray}
S^{WW}\left( x_{W}^{i}, x_{W}^{i} \right) = \frac{x_{W}^{i}}{\left( 1 -
x_{W}^{i} \right)^{2}} \left[ 3 - \frac{19}{4} x_{W}^{i} + \frac{1}{4}
\left(x_{W}^{i} \right)^{2} \right] \nonumber \\
+ \frac{2 x_{W}^{i} \ln\left(
x_{W}^{i} \right)}{\left( 1 - x_{W}^{i} \right)^{2}} \left[ 1 -
\frac{3}{4} \frac{\left( x_{W}^{i} \right)^{2}}{\left( 1 - x_{W}^{i}
\right)} \right].
\label{SWWeq}
\end{eqnarray}
If $i \neq j$:
\begin{equation}
S^{HH}\left(x_{H}^{i}, x_{H}^{j}, x_{H}^{W} \right) = \frac{x_{H}^{i}
x_{H}^{j}}{x_{H}^W} \left[ \frac{J\left(x_{H}^{i}\right) -
J\left(x_{H}^{j} \right)}{x_{H}^{i} - x_{H}^{j}} \right]
\label{SHH}
\end{equation}
with
\begin{equation}
J\left( x_{H}^{i} \right) = \frac{1}{\left( 1 - x_{H}^{i}\right)} +
\frac{\left( x_{H}^{i} \right)^{2} \ln\left( x_{H}^{i} \right)}{\left(1 -
x_{H}^{i} \right)^{2}}.
\label{J}
\end{equation}
If $ i = j$:
\begin{equation}
S^{HH}\left(x_{H}^{i}, x_{H}^{i}, x_{H}^{W} \right) =
\frac{\left(x_{H}^{i}\right)^{2}}{x_{H}^{W}} \left[\frac{1 -
\left(x_{H}^{i} \right)^{2} + 2 x_{H}^{i} \ln\left(x_{H}^{i}
\right)}{\left( 1 - x_{H}^{i} \right)^{3}} \right].
\label{SHHeq}
\end{equation}
For $ i \neq j$:
\begin{eqnarray}
S^{HW} \left( x_{W}^{i}, x_{W}^{j}, x_{H}^{i}, x_{H}^{j}, x_{H}^{W}
\right) = \frac{x_{H}^{i} x_{H}^{j}}{\left( x_{H}^{W} - 1
\right) \left(x_{H}^{i} - 1 \right) \left( x_{H}^{j} - 1 \right)} \left[ 1
- \frac{1}{8 x_{H}^{W}} \right] \nonumber \\
+ \frac{x_{H}^{i} x_{H}^{j} x_{H}^{W}}{\left( x_{H}^{W} -
1\right) \left(x_{H}^{i} - x_{H}^{W} \right) \left( x_{H}^{j} - x_{H}^{W}
\right)} \left[ \frac{3}{4} \ln\left(x_{H}^{W} \right) - \frac{7}{8}
\right] \nonumber \\
+ \frac{\left(x_{H}^{i}\right)^{2}
x_{H}^{j}}{\left(x_{H}^{i} -
x_{H}^{W} \right) \left(x_{H}^{i} - x_{H}^{j} \right) \left( x_{H}^{i} - 1
\right)} \left[ \ln\left( x_{H}^{i}\right) \left(1 - \frac{1}{4} x_{W}^{i}
\right) + \left(\frac{1}{8} x_{W}^{i} - 1\right) \right] \nonumber \\
+ \frac{\left(x_{H}^{j}\right)^{2}
x_{H}^{i}}{\left(x_{H}^{j} -
x_{H}^{W} \right) \left(x_{H}^{j} - x_{H}^{i} \right) \left( x_{H}^{j} - 1
\right)} \left[ \ln\left( x_{H}^{j}\right) \left(1 - \frac{1}{4} x_{W}^{j}
\right) + \left(\frac{1}{8} x_{W}^{j} - 1\right) \right].
\label{SHW}
\end{eqnarray}
For $ i = j$:
\begin{eqnarray}
S^{HW} \left( x_{W}^{i},x_{W}^{i},x_{H}^{i},x_{H}^{i},x_{H}^{W}
\right) = \left( x_{H}^{i} \right)^{2} \left[ \frac{ \ln \left(x_{H}^{i}
\right)}{\left( x_{H}^{W} - 1 \right) \left( x_{H}^{i} - 1 \right)^{2}}
\left( 1 - \frac{1}{4 x_{H}^{W}} \right)  \right. \nonumber \\
\left. - \frac{3}{4} \frac{x_{H}^{W} \ln \left( x_{W}^{i} \right)}{
\left(x_{H}^{W} - 1 \right) \left( x_{H}^{i} - x_{H}^{W}\right)^{2}} +
\frac{1}{
\left( x_{H}^{i} - 1 \right) \left( x_{H}^{i} - x_{H}^{W} \right)} \left(
1 - \frac{1}{4} x_{W}^{i} \right) \right].
\label{SHWeq}
\end{eqnarray}

\section{Feynman rules of the charged Higgs
in the Two Higgs Doublet Model}.

The effective Lagrangian corresponding to the $H^{\pm} f \bar{f}'$ vertex is:
\begin{equation}
L = \frac{g}{2 \sqrt{2} m_{W}} \left[ H^{+} V_{f f'} \bar{u}_{f} \left(
A + B \gamma^{5} \right) v_{\bar{f}'} + h.c. \right]
\label{L}
\end{equation}
where
$A \equiv \left( m_{f'} \tan \beta + m_{f} cot \beta \right)$
and $B \equiv \left( m_{f'} \tan \beta - m_{f} cot \beta \right)$,
$f =$ fermion (quark or lepton) and $ \bar{f'} =$ antifermion (antiquark or
antilepton). $V_{f f'}$ is  an element of the CKM matrix.

The charged-Higgs propagator is: $i / \left( K^{2} - m_{H}^{2} + i
\varepsilon \right)$.

\newpage

\end{document}